\documentclass[a4paper,12pt]{article}
\pdfoutput=1
\usepackage{graphicx,rotating,colortbl}
\usepackage{hyperref,slashed,amsfonts}
\usepackage{color}
\usepackage{amsmath}
\usepackage{mathtools}
\hypersetup{colorlinks,bookmarksopen,bookmarksnumbered,
linkcolor=blus,pdfstartview=FitH,urlcolor=rossos,citecolor=verde}

\def\circa#1{\,\raise.3ex\hbox{$#1$\kern-.75em\lower1ex\hbox{$\sim$}}\,}

\usepackage{mathtools}
\usepackage{tikzfeynman}
\usepackage{youngtab}
\usepackage{multicol}
\usepackage{color}
\definecolor{rosso}{cmyk}{0,1,1,0.4}
\definecolor{rossos}{cmyk}{0,1,1,0.55}
\definecolor{rossoc}{cmyk}{0,1,1,0.2}
\definecolor{blu}{cmyk}{1,1,0,0.3}
\definecolor{blus}{cmyk}{1,1,0,0.6}
\definecolor{bluc}{cmyk}{1,1,0,0.1}
\definecolor{verde}{cmyk}{0.92,0,0.59,0.25}
\definecolor{verdec}{cmyk}{0.92,0,0.59,0.15}
\definecolor{verdes}{cmyk}{0.92,0,0.59,0.4}

\newcommand{\Q}{\Psi}

\def\circa#1{\,\raise.3ex\hbox{$#1$\kern-.75em\lower1ex\hbox{$\sim$}}\,}

\newcommand{\beq}{\begin{equation}}
\newcommand{\eeq}{\end{equation}}

\newcommand{\bea}{\begin{eqnarray}}
\newcommand{\eea}{\end{eqnarray}}
\newcommand{\be}{\begin{equation}}
\newcommand{\ee}{\end{equation}}
\font\tenrsfs=rsfs10 at 12pt
\font\sevenrsfs=rsfs7
\font\fiversfs=rsfs5
\newfam\rsfsfam
\textfont\rsfsfam=\tenrsfs
\scriptfont\rsfsfam=\sevenrsfs
\scriptscriptfont\rsfsfam=\fiversfs
\def\mathscr#1{{\fam\rsfsfam\relax#1}}

\def\Lag{\mathscr{L}}

\def\circa#1{\,\raise.3ex\hbox{$#1$\kern-.75em\lower1ex\hbox{$\sim$}}\,}
\makeatletter

\def\hhref#1{\href{http://arxiv.org/abs/#1}{arXiv:#1}} 
\def\arXiv#1{\href{http://arxiv.org/abs/#1}{arXiv:#1}} 

\def\art{\@ifnextchar[{\eart}{\oart}}
\def\eart[#1]#2#3#4#5#6{{\rm #2}, {\em #3 \bf #4} {\rm (#6) #5} ({\em #1})}
\def\hepart[#1]#2{{\rm #2, \hhref{#1}}}
\newcommand{\oart}[5]{{\rm #1}, {\em #2 \bf #3} {\rm (#5) #4}}

\newcounter{alphaequation}[equation]
\def\thealphaequation{\theequation\hbox to
0.6em{\hfil\alph{alphaequation}\hfil}}
\def\eqnsystem#1{
\def\@eqnnum{{\rm (\thealphaequation)}}
\def\@@eqncr{\let\@tempa\relax \ifcase\@eqcnt \def\@tempa{& & &} \or
  \def\@tempa{& &}\or \def\@tempa{&}\fi\@tempa
  \if@eqnsw\@eqnnum\refstepcounter{alphaequation}\fi
\global\@eqnswtrue\global\@eqcnt=0\cr}
\refstepcounter{equation} \let\@currentlabel\theequation \def\@tempb{#1}
\ifx\@tempb\empty\else\label{#1}\fi
\refstepcounter{alphaequation}
\let\@currentlabel\thealphaequation
\global\@eqnswtrue\global\@eqcnt=0 \tabskip\@centering\let\\=\@eqncr
$$\halign to \displaywidth\bgroup \@eqnsel\hskip\@centering
$\displaystyle\tabskip\z@{##}$&\global\@eqcnt\@ne
\hskip2\arraycolsep\hfil${##}$\hfil& \global\@eqcnt\tw@\hskip2\arraycolsep
$\displaystyle\tabskip\z@{##}$\hfil
\tabskip\@centering&\llap{##}\tabskip\z@\cr}
\def\endeqnsystem{\@@eqncr\egroup$$\global\@ignoretrue} \makeatother

\newcommand{\SO}{\,{\rm SO}}

\begin{document}
\phantom{ghost}

\vspace{2cm}

\begin{center}
{\Large \bf \color{rossos}
The Half-composite Two Higgs Doublet Model\\ and the Relaxion
}\\[1.5cm]

{\bf Oleg Antipin\footnote{oleg.antipin@fi.infn.it}, Michele Redi\footnote{michele.redi@fi.infn.it}}  
\\[7mm]

{\it INFN, Sezione di Firenze, Via G. Sansone, 1; I-50019 Sesto Fiorentino, Italy}
\vspace{1.5cm}\\
{\large\bf\color{blus} Abstract}
\begin{quote}
We study a new confining gauge theory with fermions in a vectorial representation under the SM gauge group that allows for 
Yukawa interactions with the Higgs. If the fermion masses are smaller than the confinement scale this realizes a type I two Higgs doublet model 
where a composite Higgs mixes with the elementary Higgs. This class of models interpolates between an elementary and a composite Higgs 
and has interesting phenomenology with potentially observable effects in collider physics, EDMs and SM couplings but very weak bounds from indirect searches.
The very same framework can be used to realize the cosmological relaxation of the electro-weak scale recently discussed in the literature. 
\end{quote}

\thispagestyle{empty}
\end{center}
\begin{quote}
{\large\noindent\color{blus} 
}

\end{quote}
\vspace{-1.5cm}

\newpage
\tableofcontents

\setcounter{footnote}{0}

\section{Introduction}

A very simple extension of the Standard Model (SM) that is not disfavoured by experimental data and could be visible in future experiments  
is provided by a new confining gauge theory with fermions in a real representation of the  SM. These scenarios, also known as vector-like confinement \cite{sundrum}, could be relevant for LHC phenomenology and dark matter \cite{ACDM}. Moreover, while they do not address the naturalness of the  electro-weak scale in  the standard way, they could play a role in alternative explanations of the electro-weak scale \cite{relaxion,Antipin:2014qva}. 

The theories under consideration are described by the renormalizable lagrangian,
\beq 
\Lag = \Lag_{\rm SM}    +  \bar\Q_i( i\slashed{D}  - m_i ) \Q_i - \frac{1}{4g_{\rm H}^2} G_{\mu\nu}^{A} G^{A\mu\nu}
+\frac {\theta_H} {32\pi^2} G^A_{\mu\nu}\tilde{G}^A_{\mu\nu}
+ [ H\bar\Q_i(y^L_{ij}P_L+ y^R_{ij}P_R) \Q_j + \hbox{h.c.}]
\label{lagrangian1}
\eeq
where the covariant derivative contains the SM gauge fields $A_\mu$ and new $SU(N)$ fields $B_\mu$\footnote{We focus on $SU(N)$ gauge theories with matter in the fundamental representation but all the results can be generalised to other representations as well as $\SO(N)$ and ${\rm Sp}(N)$ gauge groups.}.
In this paper we will  be interested in models  where the quantum numbers of the fermions allow for Yukawa interactions with the elementary Higgs doublet $H$. 
The dynamics of this theory is well known from QCD. For sufficiently small number of flavors $N_F$ the $SU(N)$ interactions confine and the pattern 
of symmetry breaking is $SU(N_F)\times SU(N_F)/SU(N_F)$ delivering light pseudo-Goldstone bosons.
Because the fermions are vectorial under the SM the strong dynamics does not break the SM gauge symmetry dynamically,
giving very weak bounds on these type of models from present data. In particular, this class of theories features 
automatic Minimal Flavor Violation avoiding all flavor bounds.

The motivation of this work is two-fold: The addition of Yukawa couplings has a profound impact on the phenomenology of these models
that has not been considered in the literature. Whenever the quantum numbers allow for Yukawa couplings 
there exists a composite pion with equal quantum number as the Higgs, a ``Kaon''.
This doublet can and will mix with the Higgs. The low energy theory is effectively described by a two Higgs doublet model (2HDM) where 
electro-weak symmetry breaking is triggered by the  elementary Higgs and a vacuum expectation value (VEV)  for the composite Higgs is induced by the mixing. This resembles ``bosonic technicolor'' models (see \cite{Galloway:2013dma} for a recent discussion),  with the crucial difference that the new sector in isolation
does not break the electro-weak symmetry\footnote{The same theory was considered in the original work on the composite Higgs  \cite{kaplangeorgi} 
where however the elementary Higgs was used to generate the necessary vacuum of misalignment.}.
Since the SM fermions couple to the elementary Higgs in the fundamental action this effectively generates s 2HDM of type I.
We call this the Half-composite Two Higgs Doublet Model. Due to the mixing, the recently discovered Higgs particle is 
a mixture of the elementary and composite doublets so that some deviations from SM predictions for precision tests and
Higgs couplings are predicted. Yukawa couplings also introduce new CP violating phases in theory that leads to Electric-Dipole-Moments (EDMs)
for SM particles that could be the most promising signature of these models.

Secondly, theories such us (\ref{lagrangian1}) have also been considered recently in the context of a cosmological relaxation of the electro-weak scale \cite{relaxion}. In this paper we explore a slight modification of the scenario discussed in that paper with all the fermion masses smaller than the confinement scale such that these fermions participate to the chiral dynamics. This leads to a completely different phenomenology from \cite{relaxion} as the lightest particles are now scalar bound states made of fermions, in particular the composite doublet. Moreover the parameter space that allows to realize the relaxation of the electro-weak scale is different allowing a compositeness scale around TeV, for example. This is compatible with present searches and accessible to future experiments.

In what follows we work out in some detail the simplest model that allows for Yukawa couplings, the $L+N$ model: 
Dirac fermions belonging to  the fundamental rep of  $SU(N)$ with quantum numbers of a lepton doublet $L$ and a singlet $N$ under the SM.
The strong dynamics is identical to QCD with 3 flavors so that one can use the guidance of data to draw quantitative conclusions.
We will focus on the lightest states given by the pion octet. After describing the relevant low energy lagrangian we connect
to 2HDM providing estimates for the electro-weak precision parameters and Higgs couplings. Particularly important are the anomalies with SM gauge fields 
that control the decay of pion triplets and singlets similarly to $\pi_0\to \gamma\gamma$. 
They  also generate dipole interactions for SM particles, leading in  particular to significant effects for EDMs.
In section \ref{sec:relaxion} we study how the relaxation of the electro-weak scale can be realised within the low energy effective lagrangian
of our 2HDM. The mechanism is identical to \cite{relaxion} but now a crucial role is played by the composite doublet mixing
with the elementary Higgs. 

\section{Half-Composite Two Higgs doublet model}
\label{LN}

For concreteness we consider the $L+N$ model also discussed in \cite{Antipin:2014qva,ACDM,relaxion}. 
We emphasize however that our arguments work in general whenever the quantum numbers of the vectorial fermions allow for Yukawa couplings with the Higgs. 
We add to the SM  a doublet $L$ and a singlet $N$ in the fundamental representation of $SU(N)$ plus their conjugates. 
The action contains the mass terms,
\begin{eqnarray}
\mathcal{L}_M&=& m_L L L^c+ m_N N N^c+ y H L N^c+ \tilde{y} H^\dagger L^c N +h.c.\nonumber \\
&=&\frac{H}{2}\left( A\bar{\Q}_1 \Q_2 - B \bar{\Q}_1 \gamma_5 \Q_2  + h.c.\right) + m_L \bar{\Q}_2 \Q_2+ m_N \bar{\Q}_1 \Q_1
\end{eqnarray}
where $m_L$ and $m_N$ are taken to be real and we assume that they are smaller than the confinement scale of the theory.  
In the second line we introduced the Dirac notation $\Psi_1=(N,\bar{N}^c)^T$ and $\Psi_2=(L,\bar{L}^c)^T$ with $A\equiv  (y+\tilde{y}^*)$ and $B\equiv  (y-\tilde{y}^*)$. 

The action contains one CP violating phase corresponding to the imaginary part of  $m_L m_N y^* \tilde{y}^*$. 
The renormalizable lagrangian (\ref{lagrangian1}) also contains the CP violating topological $\theta_H$-term of $SU(N)$ gauge fields. 
As usual $\theta_H$ can be rotated to the fermion  mass matrix by means of a chiral transformation of the fermion fields.  The appropriate rotation depends 
on the masses, which we discuss in the appendix. For example for the hierarchy of scales $m_N\ll m_L$  it is convenient to eliminate $\theta_H$ with 
a chiral rotation on the singlet. 
This amounts to,
\begin{equation}
m_N \to m_N e^{i \theta_H}\,,~~~~~~~~~~~\tilde{y} \to \tilde{y} e^{i \theta_H}
\label{chiralN}
\end{equation}
In the opposite limit $m_L\ll m_N$ instead,
\begin{equation}
m_L \to m_L e^{i \frac{\theta_H}2}\,,~~~~~~~~~~~y \to y e^{i \frac{\theta_H}2}
\label{chiralL}
\end{equation}

The dynamics of this theory is as in QCD. Below the confinement scale $m_\rho$ a chiral condensate forms that breaks the global symmetry $SU(3)\times SU(3)\to SU(3)$ 
producing eight Goldstone bosons in the octet of $SU(3)$. Under the $SU(2)_L\times U(1)_Y$ they decompose as  $8=3_0\oplus2_{\pm 1/2}\oplus 1_0$.
The pions, including the $\eta'$ associated to the anomalous $U(1)$ axial current of the $SU(N)$ gauge theory, can be represented as,
\begin{equation}
\Pi = \left( \begin{array}{ccc}
\pi_3^0/\sqrt{2}+\eta/\sqrt{6}& \pi_3^+ & K^{+}_2 \\
\pi_3^{-} & -\pi_3^0/\sqrt{2}+\eta/\sqrt{6} & K^{0}_2 \\
K^{-}_2 & \bar{K}^{0}_2 & -2\eta/\sqrt{6} \end{array} \right) + \frac{\eta'}{\sqrt{3}} 1\hspace{-0.9ex}1_3.
\label{QCDpions}
\end{equation}

The chiral lagrangian including anomalies reads, 
\begin{eqnarray}
\mathcal{L}&=&\frac{f_\pi^2}{4}{\rm Tr} [D_\mu U D^\mu U^\dagger]+  (g_\rho f_\pi^3 {\rm Tr}[MU]+ h.c) +\frac{f_\pi^2}{16}\frac a N \bigg[ \ln (\hbox{det}\ U)- \ln (\hbox{det}\ U^\dagger)\bigg]^2 \nonumber \\
&-&\frac{N}{16\pi^2 f_\pi}\sum_{G_1,G_2}g_{G_1}g_{G_2} {\rm Tr}[\pi^a T^a F^{(G_1)} \tilde{F}^{(G_2)}]+ 
 \frac {3 g_2^2 g_\rho^2 f_\pi^4 } {2(4\pi)^2} \sum_{i=1..3} {\rm Tr}[U T^i U^\dagger T^i]
\label{lagrangian}
\end{eqnarray}
where, 
\be
M = \left( \begin{array}{ccc}
m_L & 0 & y h^+ \\
0 & m_L & y h^0 \\
\tilde{y}  h^{-} & \tilde{y}  h^{0\dagger}  & m_N \\  
\end{array} \right)\, \qquad \quad {\rm and} \qquad \quad U\equiv e^{i \sqrt{2}\Pi/f_\pi}
\label{massmatrix}
\ee
The covariant derivative takes the form $D_\mu U=\partial_\mu U -i  A_\mu U+i U  A_\mu$ where $A_\mu$ are the SM gauge fields. 
The coupling scales as $g_\rho \sim 4\pi/\sqrt{N}$ and for $N=3$ its value is around 7 from QCD data. The cut-off of the chiral lagrangian can be taken to be the mass of the lightest vector resonance that scales as $m_\rho \sim g_\rho f_\pi$ corresponding to the mass of the lightest vector resonance. The last term on the first line describes the effect of the $U(1)$ axial anomaly of $SU(N)$ that reproduces the mass of the singlet $\eta'$, $m_{\eta'}^2= 3 a/N+ {\cal O}(M)$ and accounts for the proper selection of the vacuum. This is important in the region of small fermion masses where the vacuum is displaced from the origin.  In what follows we will consider the region with non-zero fermion masses where expansion around the origin is appropriate.  On the second line we have included the 1-loop gauge contribution (an analogous term arises from the Yukawa couplings) and also schematically the effect of anomalies with SM gauge fields. The full non-linear Wess-Zumino-Witten lagrangian will not be needed for the present analysis.  

Expanding the potential in  (\ref{lagrangian})  to cubic order in the fields around $\Pi=0$ we find: 
\footnote{Mass terms are in general complex when $\theta_H$ is different from zero. This generates a tadpole for $\eta$ and CP violating processes such as $\eta \to \pi \pi$.  The tadpole for $\eta$ can be eliminated by rephrasing the fields so that ${\rm Im}[m_L-m_N]=0$ which is equivalent to the Dashen's equations for the vacuum discussed in the appendix. For $m_N \ll m_L$ this gives (\ref{chiralN}) and in the opposite limit (\ref{chiralL}).}
\begin{eqnarray}
\label{expanded}
&&\mathcal{L}_m= g_\rho f_{\pi}^3 Tr[MU]+ h.c \ + 
 \frac {3 g_2^2 g_\rho^2 f_\pi^4 } {2(4\pi)^2} \sum_{i=1..3} {\rm Tr}[U T^i U^\dagger T^i] \nonumber \\
&\approx&{\rm Re}[4 m_L+ 2 m_N] g_\rho f_\pi^3-m_{K_2}^2 K_{2}^{\dagger } K_2  -\frac {m_{\pi_3}^2}2 \pi_3^a\pi_3^a- \frac{m_{\eta}^2}{2}\eta^2 \nonumber \\
&-&i\sqrt{2} g_\rho f_\pi^2 B K_{2 }^{\dagger} H -\frac{g_\rho}{\sqrt{2}} Af_\pi \left(K_{2}^{\dagger} \sigma^a\pi^{a}_3 -\frac{\eta K_{2}^{\dagger}}{\sqrt{3}} \right) H +h.c. \nonumber \\ 
&+&\frac{g_\rho ({\rm Im}(m_L)- {\rm Im}(m_N)) \eta}{\sqrt{3}} \left(4 f_{\pi}^2-\frac{2\eta^2}{9}\right)+\frac{ 2g_\rho\eta}{\sqrt{3}}\left(K_{2}^{\dagger } K_2 {\rm Im}(m_N) - \pi_3^a \pi_3^a {\rm Im}(m_L)\right)\nonumber \\
&+& \frac{2 }{3}g_\rho(2{\rm Im}(m_L)+ {\rm Im}(m_N))K_2^{\dagger} \sigma^a K_2 \pi_3^a
\end{eqnarray}
where we have defined the elementary Higgs doublet $H= (h^+ ,  h^0)^T$ 
the composite singlet "$\eta$", the composite "Kaon" doublet $K_2= (K^+_2 ,  K^0_2)^T$ and the composite "Pion" triplet $\pi_3^a$ .

The pion mass parameters are given by\footnote{We neglect the loop contribution from Yukawa couplings discussed in appendix of \cite{Antipin:2014qva}
as we will assume $y, \tilde{y}\ll1$ in what follows.  For the gauge contribution rescaling the electromagnetic splitting of pions in QCD we estimate $\Delta m^2_g=\frac{3 (g_2 m_{\rho})^2}{(4\pi)^2} J(J+1)$ where $J$ is the isospin of the multiplet.},
\begin{eqnarray}
\label{pionmasses}
m^2_{\pi_3}&\approx& \frac{6 g_2^2 g_\rho^2 }{(4\pi)^2}f_\pi^2+ 4 {\rm Re}[m_L]g_{\rho} f_\pi\\
m^2_{K_2}&\approx&  \frac{9 g_2^2 g_\rho^2}{4(4\pi)^2}f_\pi^2+ 2{\rm Re}[m_L+m_N]g_{\rho} f_\pi\nonumber\\
m_{\eta}^2&\approx& \frac{4}{3}{\rm Re}[m_L+2m_N]g_{\rho} f_\pi \ . \nonumber
\end{eqnarray}
Scalars charged under the SM acquire a positive mass from gauge loops.
The singlet mass is only controlled by the elementary fermion masses and could be very small.
In the limit of massless fermions the Yukawa interactions induces a small mass \cite{Antipin:2014qva},
\begin{equation}
m_\eta^2 \sim |y \tilde{y}|\frac {g_\rho^2 f _\pi^2}{m_{K_2}^2} v^2
\label{masseta}
\end{equation}
proportional to the Higgs VEV $v= 246$ GeV. 

The strong sector lagrangian should be supplemented with the usual SM Higgs lagrangian
\be
\mathcal{L}_{}=\left|D_\mu H\right|^2 -\lambda (H^\dagger H)^2+ m^2 H^\dagger H
\ee
where the parameters must be chosen to reproduce the correct electro-weak vacuum and Higgs mass.

The truncated lagrangian above can be used as long as the VEVs 
of the fields are much smaller than $f_\pi$ which will always be the case in the region of parameters space studied in this paper. 

\section{Phenomenology}

For $m_{L,N}< m_\rho$, apart from the pseudo-Goldstone bosons, the spectrum of our $L+N$ model contains  heavy  mesonic and baryonic states. 
For the collider phenomenology of mesonic spin-1 resonances in the framework of vector-like confinement we refer to \cite{sundrum}.
Baryonic DM candidates in the framework of vector-like confinement were studied in detail in the context of composite Dark Matter \cite{ACDM}. 
In the regime of the $L+N$ model under consideration, while lightest baryons can be neutral, they cannot be the dominant DM component due to tree level coupling to the $Z$; 
they should either decay through higher dimension operators or provide a suppressed relic density. 
In other models the baryons could play the role of DM. For example, the 5-flavors $V+L$ model where we replace the singlet $N$ with a triplet $V$, 
also allows for Yukawa couplings with the elementary Higgs and contains a baryon DM candidate $VVV$ transforming as a weak triplet \cite{ACDM}. 

We will focus here on the phenomenology associated to pions that are the lightest particles beyond the SM. A lower bound on their mass exists due to gauge interactions which are strictly positive. From eq. (\ref{pionmasses}),
the doublet acquires a mass of order $m_\rho/10$ that we take as the lower limit.

Most important for the phenomenology is the term $K_2^\dagger H$ on the second line of eq. (\ref{expanded}) that induces a mixing between the elementary and 
the composite Higgs,
\begin{equation}
K_2\approx  \epsilon H\,,~~~~~~~~~~~~~~~~~~~~~~~ \epsilon\equiv i \sqrt{2} (y-\tilde{y}^*)\frac {g_\rho f_\pi^2}{m_{K_2}^2}
\end{equation}
Due to this mixing the SM Higgs mass eigenstate has a component of the composite  doublet leading to modification of SM couplings 
and exotic decays of pions. Therefore, the model interpolates between an elementary and a composite Higgs.

Note that $\epsilon$ could be greater than one with perturbative Yukawa couplings. 
In this regime the electro-weak symmetry is mostly broken by the composite Higgs. 
In this paper we will only study the small mixing region where experimental bounds are weaker.
A general study will appear elsewhere.

From the lagrangian (\ref{expanded}) we obtain the VEVs for the heavy partners,
\begin{eqnarray}
\langle K^0_2 \rangle &=&\sqrt{2}\frac{i B g_\rho f_\pi^2 }{m_{K_2}^2} \langle h^0 \rangle \equiv  \epsilon \langle h^0 \rangle   \qquad  \\
 \langle \pi_3^0 \rangle &=&\frac{g_\rho f_\pi  \langle h^0 \rangle }{\sqrt{2} m_{\pi_3}^2} \left[A \langle K_2^{0 *} \rangle +A^* \langle K_2^{0 } \rangle\right]=-\frac{4{\rm Im}(y \tilde{y}) g_\rho^2 f_\pi^3}{m_{K_2}^2 m_{\pi_3}^2} \langle h^0 \rangle^2\nonumber \\
 \langle \eta \rangle& =& \frac{\langle \pi_3^0 \rangle}{\sqrt{3}}\frac{m_{\pi_3}^2}{m_{\eta}^2}\nonumber \ ,
 \label{VEVs}
\end{eqnarray}
where the VEV for the neutral component of the elementary Higgs doublet $\langle h^0 \rangle $  is given by
\be
 \langle h^0 \rangle^2 = \frac{m^2+|\epsilon|^2 m_{K_2}^2}{2\lambda} \ .
\ee

\subsection{Standard Model Couplings}

Since the lighter Higgs doublet is a mixture of an elementary and a composite Higgs it inherits some properties of the scenarios where 
the Higgs is a composite pseudo-Goldstone boson. In conventional composite Higgs models the effects of compositeness can be parametrized in terms of effective operators \cite{SILH} and the same can be done here.  Let us first discuss electro-weak precision tests. To obtain the new physics contribution to the effective action for the light Higgs 
we need to dress the higher order terms for the composite  pions with the mixing. One finds,
\begin{equation}
\hat{T} \sim \frac {v^2} {f_\pi^2} |\epsilon|^4\,,~~~~~~~~~~~~~~~~~~\hat{S}\sim  \frac{m_W^2}{m_\rho^2} |\epsilon|^2
\end{equation}
The correction to $\hat{T}$ originates from the fact that the strong dynamics does not respect custodial symmetry and could be absent in other models.
The effects are doubly suppressed by the mixing and the compositeness scale. The experimental bound on $\hat{T}$ and $\hat{S}$ are around 
$10^{-3}$ implying,
\begin{equation}
\epsilon < 0.2 \times \sqrt{\frac{f_\pi} {v}}\,~~~~~~~~~~~{\rm or}~~~~~~~~~~~~\epsilon < 0.03 \times \frac {g_\rho}{g_2}\times \frac{f_\pi} {v}
\end{equation}

Concerning Higgs couplings, we have effectively a type-I two Higgs doublet model with\footnote{The mixing with the triplet is not relevant to leading order.}:
\be
 \langle K_2 \rangle=\left( \begin{array}{cc}
0   \\
 \frac{v_2 e^{i \rho}}{\sqrt{2}}  \\  
\end{array} \right)  \\  \quad 
\langle H \rangle=\left( \begin{array}{cc}
0   \\
 \frac{v_1 }{\sqrt{2}}  \\  
\end{array} \right) \\ \quad  \langle h^0 \rangle =\frac{v_1}{\sqrt{2}} \qquad \frac{v_2 e^{i \rho}}{\sqrt{2}} \equiv   \epsilon \langle h^0 \rangle \qquad {\rm tan}\beta= \frac{v_2}{v_1}= |\epsilon| \quad 
\ee
and $ v_1^2+v_2^2 =v_{}^2 = (246 \ {\rm GeV})^2$.

In order to identify the physical degrees of freedom, it is convenient to introduce transformation :
\be
H= \cos\beta\  \Phi_1-\sin\beta \ \Phi_2 \qquad K_2= (\sin\beta\ \Phi_1+ \cos\beta \ \Phi_2) e^{i \rho}
\label{mixLO}
\ee
which takes the two doublets ($H$ and $K_2$) to the eigenstates $\Phi_1$ and $\Phi_2$ such that $\langle \Phi^0_1 \rangle= v_{}$ and $\langle \Phi^0_2 \rangle=0$:
\be
 \Phi_1=\left( \begin{array}{cc}
G^+   \\
\frac{v_{}+h_1+i G^0}{\sqrt{2}}  \\  
\end{array} \right)  \\  \quad 
 \Phi_2=\left( \begin{array}{cc}
H^+   \\
 \frac{h_2+i A_0}{\sqrt{2}}  \\  
\end{array} \right) \\ 
\ee
where $G^{\pm}$ and $G^0$ are Goldstone bosons eaten by $W$ and $Z$. We have eliminated the phase $\rho$ 
from the quadratic lagrangian that will appear in the trilinear term $K_{2}^{\dagger} \pi_{3} H$ leading to CP violating effects.

For the spectrum of CP-even states $h_{1,2}$ and CP-odd state $A_0$ we obtain:
\be
m_{A_0}^2 = m_{H^{\pm}}^2= m_{K_2}^2 (1+|\epsilon|^2)\ , \ \  
M_0^2= \left( \begin{array}{cc}
m_{h}^2 & -|\epsilon| m_{h}^2  \\
 -|\epsilon| m_{h}^2& m_{K_2}^2\\  
\end{array} \right) 
\ee
where $m_h^2= 2 m^2=2 \lambda v^2$. The mixing angle between  the neutral states $h_{1,2}$ is:
\be
\tan 2\delta\approx -\frac{2 |\epsilon| m_{h}^2}{m_{h}^2-m_{K_2}^2} 
\ee
and we identify the lightest mass eigenstate $\tilde{h}_1$ with the discovered 125 GeV Higgs boson while the $\tilde{h}_2$ is the heavy CP-even scalar.
The coupling of the Higgs to the SM vector bosons $V$ is (see \cite{wells} for a review):
\be
\frac{\tilde{h}_1 VV}{(hVV)^{SM}}=\frac{\cos (\beta-\delta)\langle h^0 \rangle+\sin (\beta-\delta)\langle K^0_2 \rangle}{v_{}} = \cos \delta \approx 1-\frac{|\epsilon|^2 m_h^4}{2(m_h^2-m_{K_2}^2)^2} 
\ee
while the coupling to the SM fermions is:
\begin{eqnarray}
&&y_f \frac{v_1}{\sqrt{2}}\left(1+\frac{h^0}{v_1}\right) \bar{f}f = y_f \frac{v_{} \cos \beta}{\sqrt{2}}\left(1+\frac{\tilde{h}_1\cos (\beta-\delta) }{v_{}\cos \beta}\right) \bar{f}f  \nonumber \\
&&\Longrightarrow \quad \frac{\tilde{h}_1\bar{f}f}{(h\bar{f}f)^{SM}}= \frac{\cos (\beta-\delta)}{\cos \beta} \approx 1+\frac {|\epsilon|^2}2 \frac{ (2 m_h^2 m_{K_2}^2-3 m_h^4)}{(m_h^2-m_{K_2}^2)^2}\,.
\end{eqnarray}
There are also small effects from compositeness. These scale as $v^2/f_\pi^2$ times four powers of the mixing and are therefore negligible
in the regime where the light Higgs is mostly elementary.

Let us note that for small fermion masses the singlet $\eta$ could be lighter than the Higgs. 
Higher order terms in eq. (\ref{expanded}) include interactions of the form $m_K^2/f_\pi^2 \eta^2 |K|^2$ that
upon mixing of $K$ with the elementary Higgs allow the decay  $H \to \eta\eta$.
We estimate:
\begin{equation}
\Gamma[h\to \eta \eta]\sim \frac 1 {8\pi} |y-\tilde{y}^{*}|^4 \frac {m_\rho^4}{m_{K_2}^4} \frac {v^2}{m_h}
\end{equation}
Comparing with the bound on the invisible width of the Higgs of around $10\%$ we find,
\begin{equation}
|y- \tilde{y}^*| \circa{<} 0.1\times  \frac {m_{K_2}}{m_\rho}
\end{equation}

\subsection{Collider Constraints}

The triplet $\pi_3$ and the doublet $K_2$  can be pair-produced  through SM gauge interactions.
In addition their production can also be mediated by composite spin-1 resonances. We will study in detail the collider 
phenomenology in a separate publication. We here only consider the first production mechanism that is model independent.

\begin{figure}[t]
\begin{center}
\includegraphics[width=0.5\textwidth]{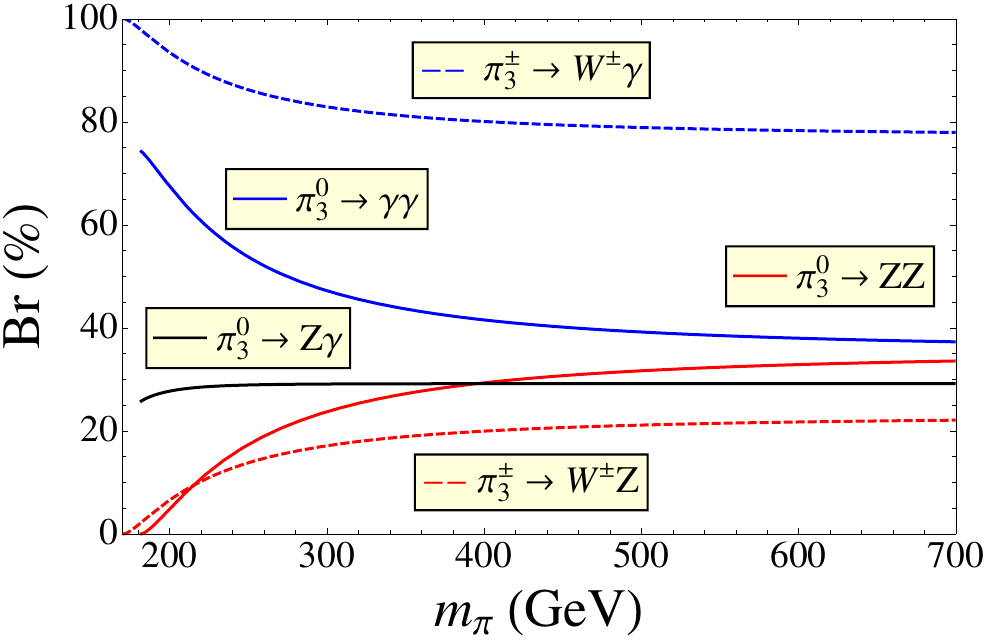}\includegraphics[width=0.51\textwidth]{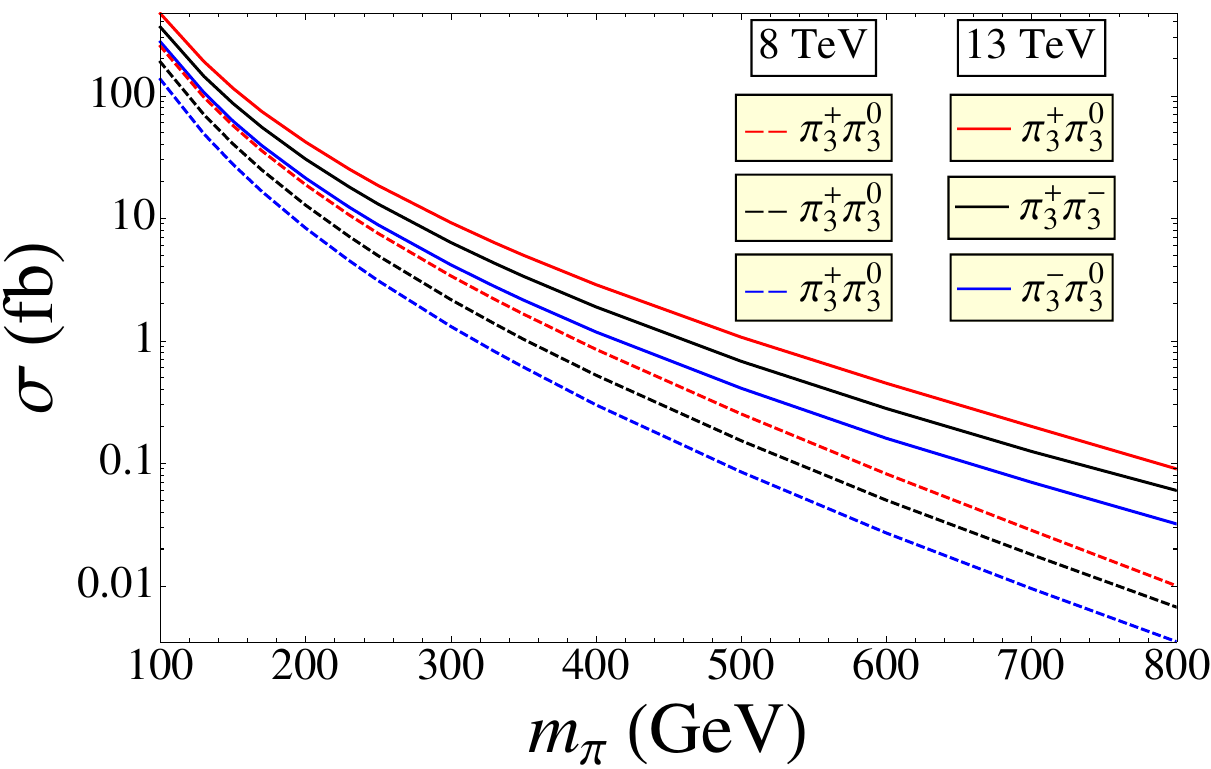}
\caption{a) On the left, branching fractions of the neutral (solid) and charged (dashed) pion triplets $\pi_3^{0,\pm}$ in $\gamma\gamma$, $ZZ$, $Z\gamma$, $W^{\pm}\gamma$ and  $W^{\pm} Z$. b) On the right, partonic production cross-section of scalar triplets at 8 TeV and 13 TeV.}
\label{Br}
\end{center}
\end{figure}

The different pions multiplets have very different fate. 
Four pion states $(\pi_3, \eta)$ decay promptly to a pair of the SM gauge bosons via a chiral anomaly:
\begin{eqnarray}
L_{\rm anomaly}^{\pi_3}&=&-\frac{g_1 g_2 N \epsilon_{\alpha\beta\mu\nu} B^{\alpha\beta} W_j^{\mu\nu}}{16 \pi^2 f_{\pi}} \pi_3^i \ {\rm Tr}[T_i T_j Y] =\frac{g_1 g_2 N   }{32 \pi^2 f_{\pi}}  \pi^a_3 W^a_{\mu \nu} \tilde{B}^{\mu \nu}\nonumber \\
&=&\frac{g_1 g_2 N}{64 \pi^2 f_{\pi}}\left[\pi_3^{\pm}\tilde{W}^{\mp \mu \nu} (\cos \theta_W F^{\mu \nu}-\sin \theta_W Z^{\mu \nu})\right . \\
&+&\left . \frac{1}{2}\sin 2\theta_W (F^{\mu \nu}\tilde{F}^{\mu \nu}-Z^{\mu \nu}\tilde{Z}^{\mu \nu})\pi_3^{0} +\cos 2\theta_W Z^{\mu \nu}\tilde{F}^{\mu \nu}\pi_3^{0} \right] \nonumber
\label{tripletanomaly}
\end{eqnarray}
where we used the fact that the hypercharge generator is $ Y =-\frac{1}{2}${\rm Diag}[1,1,0] for an $L+N$ model.
For the $\eta$ we obtain :
\begin{eqnarray}
&&L_{\rm anomaly}^{\eta}=
-\frac{ N }{32 \sqrt{3} \pi^2 f_{\pi}}\eta \left[ g_2^2 \tilde{W}^{i\mu\nu} W^i_{\mu\nu} +g_1^2 \tilde{B}^{\mu\nu} B^{\mu\nu}\right]\ \nonumber \\
&=&-\frac{ N \eta}{32 \sqrt{3} \pi^2 f_{\pi} (g_2^2+g_1^2)}  \left[(g_1^4+g_2^4) \tilde{Z}^{\mu\nu} Z^{\mu\nu}+2 g_1^2 g_2^2\tilde{F}^{\mu\nu} F^{\mu\nu}+2 g_1 g_2(g_2^2-g_1^2) \tilde{Z}^{\mu\nu} F^{\mu\nu}\right. \nonumber \\
&+&\left . g_2^2(g_1^2+g_2^2) \tilde{W}^{+\mu\nu} W^{-\mu\nu} \right] \ .\nonumber
\label{singletanomaly}
\end{eqnarray}
Since the singlet $\eta$ does not have SM interactions at tree-level, its production is suppressed and we concentrate on the phenomenology of the triplet $\pi_3$. The branching fractions are plotted in (left) Fig. \ref{Br} for the decay of the neutral and charged pion triplets $\pi_3^{0,\pm}$. 

From the Fig.\ref{Br} the dominant decay mode for the charged state is $\pi_3^{\pm} \to W^{\pm}  \gamma$ while for the neutral one it is $\pi_3^{0} \to \gamma  \gamma$. Therefore, the most interesting signal arises from the s-channel production of the $\pi_3^{\pm}  \pi_3^{0} $ via intermediate $W^{\pm}$ with the subsequent decay into $3 \gamma \ + \ W^{\pm}$ \cite{Kilic:2010et,Freitas:2010ht}:
\be
pp\to W^{\pm} \to \pi_3^{\pm}  \pi_3^{0} \to 3 \gamma \ + \ W^{\pm} \ .
\ee

An experimental bound on such decays can be extracted from the search for fermiophobic Higgs bosons in the $3\gamma\ +\ X$ final state by CDF experiment  \cite{CDF}. The exclusion translates into upper bound on the production cross-section of fermiophobic Higgs $\sigma\circa{<}$ 3 fb at the 95 \%\ C.L. Using the production cross-sections for $\pi_3^{\pm}  \pi_3^{0} $ at Tevatron computed in  \cite{Freitas:2010ht}, we extract a bound on the mass of the triplet $m_{\pi}\circa{>} 230$ GeV. The same study has not yet been performed by the LHC that would have a higher reach. Let us briefly discuss the doublet $K_2$ that is produced with cross-section similar to the triplet but can only decay through the emission of the Higgs. This is due to the fact that $K_2$ is made of fermions in different representations under the SM so that it carries a "species number" \cite{sundrum}  only broken by the Yukawa interactions. 
The four Kaons with species number $K_2=(K_2^{\pm} , K_2^0, \bar{K}^{0}_2)$ decay with emission of Higgs:
$K_2 \to h\pi_3,$ and $K_2 \to h\eta$ whichever is allowed by phase space and quantum numbers. 
For example, neglecting $h$ and $\eta$ masses, we obtain:
\be
\Gamma_{K_2 \to h\eta} = \frac{|A|^2}{48 \pi }\frac {m_{\rho}^2}{m_{K_2}} \ .
\ee
This leads to final state with large multiplicities of particles.

\subsection{Electron Dipole Moment}

For complex Yukawa couplings the model contains a CP violating phase given by ${\rm Arg}[m_L m_N y^* \tilde{y}^*]$. In general new CP violating phases exist whenever the quantum numbers allow  for Yukawa couplings. These phases generate EDMs for the SM particles, the electron and neutron in particular. In the weakly coupled regime this is completely analogous to split supersymmetry \cite{splitsusy2},  for the $L+N$ model the Higgsino and the bino.  When the masses are above the confinement scale  a weakly coupled computation is justified and the leading effect arises from 2-loop Barr-Zee diagrams \cite{Barr:1990vd}. For the $L+N$ model the contribution is suppressed because there are no diagrams with photons in the loop.

We here give an estimate of the electron EDM using the low energy chiral lagrangian. Since in the perturbative regime the EDM grows diminishing the masses of the  fermions, one expects the effect to be maximal when the fermions are lighter than the  confinement scale and this is indeed what we find. Upon diagonalization of the mixing between $K_2$ and $H$ in (\ref{expanded}), the relevant terms of the lagrangian read
\begin{eqnarray}
L^{\rm EDM}&=&-\frac{m_{\pi_3}^2}{2}( \pi_3^a)^2- \frac{m_{\eta}^2}{2} \eta^2+\frac{4 {\rm Im}(y \tilde{y}) g_{\rho}^2 f_\pi^3}{m_{K_2}^2}\left(H^\dagger \sigma^a H \pi^a_3- \frac 1 {\sqrt{3}} \eta H^\dagger H\right)\nonumber \\
&+&\frac{g_1 g_2 N   }{32 \pi^2 f_{\pi}}  \pi^a_3 W^a_{\mu \nu} \tilde{B}^{\mu \nu}-\frac{ N }{32 \sqrt{3} \pi^2 f_{\pi}}\eta \left[ g_2^2 \tilde{W}^{i\mu\nu} W^i_{\mu\nu} +g_1^2 {\tilde{B}}^{\mu\nu} B^{\mu\nu}\right]
\end{eqnarray}
Integrating out at tree level $\pi_3$ and $\eta$ we obtain the following terms in the effective action for the Higgs,
\begin{equation}
L^{\rm EDM} \subset  \frac N {24\pi^2}\frac{ {\rm Im}(y \tilde{y}) g_{\rho}^2 f_\pi^2}{m_{K_2}^2 m_{\pi_3}^2 m_\eta^2} \left[3 g_1 g_2 m_{\eta}^2\,H^\dagger \sigma^a H W_{\mu\nu}^a \tilde{B}_{\mu\nu}+ m_{\pi_3}^2 H^\dagger  H \left( g_2^2 \tilde{W}^{i\mu\nu} W^i_{\mu\nu} +g_1^2 {\tilde{B}}^{\mu\nu} B^{\mu\nu}\right)\right]\,.
\end{equation}
This contains the coupling to the photon,
\be
L_{\rm EDM}^{\rm eff}\subset -\frac{e^2 N}{24 \pi^2}  \frac{{\rm Im}(y \tilde{y}) (3 m_{\eta}^2-2m_{\pi_3}^2) m_{\rho}^2}{m_{\pi_3}^2 m_{\eta}^2 m_{K_2}^2 } F\tilde{F} h^{0\dagger} h^0 \equiv 
-\frac{c_H}{\Lambda^2}F\tilde{F} h^{0\dagger} h^0
\ee
where we have defined $\Lambda^2\equiv \frac{m_{\eta}^2 m_{\pi_3}^2 m_{K_2}^2}{(3 m_{\eta}^2-2m_{\pi_3}^2) m_{\rho}^2}$ and $c_H\equiv N\frac{\alpha}{6\pi}{\rm Im}(y \tilde{y})$. This operator renormalizes EDM operator of the electron generating \cite{EDM},
\be
d_e\approx \frac{e m_e c_H}{4\pi^2 \Lambda^2} \log \frac{\Lambda^2}{m_h^2}
\ee
This approximation is valid for masses of the  triplet and singlet greater than the Higgs mass. We neglect for simplicity contribution from the couplings to  $W$ and $Z$
that are typically subleading \cite{splitsusy2}.  Numerically one finds,  
\begin{equation}
d_e \approx  10^{-27}\, {\rm e\,cm} \times  {\rm Im}[y \tilde{y}] \times \frac N 3 \times \left(\frac {\rm TeV}{m_{\pi_{3},\eta}}\right)^4 \times \left( \frac {m_\rho}{\rm TeV}\right)^2
\end{equation}
to be compared with the experimental limit $d_e < 8.7 \times 10^{-29}\, {\rm e\,cm}$ at $90\%$ C.L. For $\mathcal{O}(1)$ phases and pions at the TeV scale
this is within the ballpark of current experiments. 

\section{Relaxion Mechanism}
\label{sec:relaxion}

In this section we connect to the ideas on the dynamical relaxation of the electro-weak scale introduced in  Ref. \cite{relaxion}.
The ``relaxion'' mechanism relies on the existence of an axion-like field $\phi$ that scans the Higgs mass during the early universe. 
Once the Higgs VEV develops, the back reaction is such that $\phi$ stops the scanning of the electro-weak VEV producing 
a technically natural electro-weak scale. We refer to the original work for all the details of the mechanism and only focus on the 
model building aspects here.

To realize this scenario one needs a potential for $\phi$ that grows after electro-weak symmetry breaking but does not  
generate large perturbative contributions to the $\phi$ potential. Following \cite{relaxion} this can be achieved in the $L+N$ models 
if $\phi$ has an axion-like coupling \cite{relaxion},
\begin{equation}
\frac 1 {32\pi^2} \frac {\phi}{f} \tilde{G}_{\mu\nu}G^{\mu\nu} \ .
\end{equation}
Contrary to the original reference we assume that both $m_L$ and $m_N$ are below the confinement scale $m_\rho=g_\rho f_\pi$
so that they  participate to the chiral dynamics and cannot be integrated out.  As we have seen in this regime the strong dynamics 
produces a composite Higgs doublet that mixes with the elementary Higgs.  This resembles the model in the appendix B of \cite{pomarol} with a notable difference:
in that case the second Higgs is also elementary and the couplings are chosen to respect an approximate $SU(2)_R$ symmetry.
Among the light degrees of freedom our action also features a  triplet and a singlet that however will not play an important role in what follows.

We are interested in computing the potential of $\phi$ induced by the strong dynamics. 
To  do this we just have to compute the vacuum energy as a function of $\theta_H$ and replace,
\begin{equation}
\theta_H \to \frac {\phi} {f} \ .
\end{equation}
In the confined phase  this can be done starting from the lagrangian  (\ref{expanded}) and  integrating out the heavy scalars.
To leading order only the doublet  contributes, 
\begin{equation}
K_2\approx i \sqrt{2}\frac {g_\rho f_\pi^2}{m_{K_2}^2} (y - \tilde{y}^*)H 
\end{equation}
where we rotated $\theta_H$ into the mass matrix as in eqs. (\ref{chiralN}), (\ref{chiralL}) so that $y, \tilde{y}$ contain the $\theta_H$ dependence. 
Plugging into the action we find\footnote{When $m_{K_2}$ is dominated by the fermion masses the computation of the Higgs potential can be done in a simpler fashion.
In the limit $m_L\gg m_N$ the mass matrix (\ref{massmatrix}) can be diagonalized with eigenvalues $m_-\sim m_N- \frac{y\tilde{y} v^2}{m_L}$ and $m_+\sim m_L+\frac {|y|^2 v^2+|\tilde{y}|^2 v^2}{2 m_L}$
that directly reproduces the vacuum energy.},
\begin{equation}
E(\theta_H)\approx  -{\rm Re}[4 m_L+ 2 m_N] g_\rho f_\pi^3  +  2\frac{g_\rho^2 f_\pi^4}{m_{K_2}^2}\left[|y|^2+|\tilde{y}|^2 - 2 {\rm Re}[y\tilde{y}] \right] |H|^2 + {\cal O}(H^4)
\label{relaxionpot}
\end{equation}

The potential depends on the fermion masses. Using eqs. (\ref{chiralN}),(\ref{chiralL}) one finds,
\begin{eqnarray}
m_N\ll m_L:&&E(\theta_H)\approx  - 2m_N  g_\rho f_\pi^3  \cos \theta_H+   2 \frac{g_\rho^2 f_\pi^4}{m_{K_2}^2}\left[|y|^2+|\tilde{y}|^2 - 2 |y\tilde{y}| \cos\left(\theta_H-\theta_0\right) \right] |H|^2\,, \nonumber \\
m_L\ll m_N:&&E(\theta_H)\approx  -4  m_L g_\rho f_\pi^3  \cos \frac{\theta_H}2+  2 \frac{g_\rho^2 f_\pi^4}{m_{K_2}^2}\left[|y|^2+|\tilde{y}|^2 - 2 |y\tilde{y}| \cos\left(\frac {\theta_H}2-\theta_0\right) \right] |H|^2 \nonumber \\
\label{barrier}
\end{eqnarray}
where the reference value $\theta_0$ is determined by the phases of the Yukawa. 
The general case valid for arbitrary masses can be found in the appendix.

The last term  in parenthesis in (\ref{barrier}) gives a periodic potential that grows with the Higgs VEV and acts as a barrier 
for the motion of $\phi$.  The height of the barrier is,
\begin{equation}
B_H \sim |y \tilde{y}|\frac{g_\rho^2 f_\pi^4}{m_{K_2}^2} v^2 \ .
\label{barrierH}
\end{equation}

In order for the relaxation mechanism to work the Higgs dependent contribution to the $\phi$ potential should be  dominant. 
An important contribution to the $\phi$ potential arises at 1-loop. This can be estimated from closing the Higgs loop in (\ref{barrier}).
This quadratically divergent contribution is cut off at  $m_{K_2}$ where the effective theory breaks down giving\footnote{Without integrating out the $K_2$ in  (\ref{expanded}) this contribution arises from the logarithmic divergent diagram with $K_2$ and $H$ running in the loop. The cutoff of the $\log$ can be taken $m_\rho$.},
\begin{equation}
B_0\sim \frac {|y\tilde{y}|} {16\pi^2} g_\rho^2 f_\pi^4  \ .
\end{equation}
In order for the Higgs dependent term (\ref{barrierH}) to dominate one finds,
\begin{equation}
m_{K_2}< 4\pi v \ .
\end{equation}
Curiously this relation demands the existence of new physics below $4\pi v$ just as the SM without the Higgs boson, 
but the physical origin is completely different. Compared to \cite{relaxion} the important difference is that the dynamical scale can be up to a factor 10 larger
because the loop is cutoff at $m_{K_2}$ that can be parametrically smaller than $g_\rho f_\pi$. 
For example, we may choose $m_{K_2}\sim f_\pi \sim$ 500 GeV and $m_\rho\sim$ 5 TeV which is phenomenologically safe and can be tested 
at LHC run II.

We also need to require that the Higgs independent term in (\ref{barrier}) is subdominant,
\begin{equation}
{\rm Min}[m_L,m_N]< \frac {g_\rho f_\pi}{m_{K_2}^2}v^2
\label{mincond}
\end{equation}
Various limits exist. Let us first consider the case where the gauge contribution to the mass of $K_2$ can be neglected because either $m_L$ or $m_N$ are large. For $m_N\ll m_L< m_\rho$ (corresponding to  $m_{K_2}^2\approx \frac 1 2 m_{\pi_3}^2\approx \frac 3 2  m_\eta^2\approx 2 m_L g_\rho f_\pi$) 
we get the same estimate for the barrier as in \cite{relaxion},
\begin{equation}
B_H \sim |y\tilde{y}|\frac  { g_\rho f_\pi^3}{m_L} v^2
\end{equation}
and from (\ref{mincond}) we derive
\begin{equation} 
m_N < |y\tilde{y}|\frac  {v^2}{m_L}  \ .
\end{equation}
Above the confinement scale, $m_N$ receives radiative corrections from loops 
with $H$ and  $L$,  $\delta m_N \sim y \tilde {y}/(4 \pi)^2  m_L \log \Lambda/m_\rho$. This  implies the mild bound,
\begin{equation}
m_L < \frac {4\pi v}{\sqrt{\log \Lambda/ m_\rho}} \ .\label{logmL}
\end{equation}

Analogous formulae apply for $m_L \ll m_N< m_\rho$.  There is no phenomenological problem in having the mass of the fermion doublet  small 
because the composite states acquire a mass from gauge interactions. In particular from (\ref{pionmasses})
it follows that the triplet mass is dominated by the gauge contribution. We then expect it to be the lightest state.
As we have seen the bounds on these electro-weak charged scalars is around 200-300 GeV that translate into a dynamical scale 
$m_\rho$ of a few TeV.

Finally we may consider the hierarchy $m_{L,N}\ll m_\rho$ where the pion masses are dominated by gauge loops. 
Here we find,
\begin{equation}
B_H \sim  |y\tilde{y}| \frac { f_\pi^2}{g_2^2} v^2 \ .
\end{equation}
The bound (\ref{logmL}) is automatically satisfied in this case.

It is interesting to consider the limit $m_{L,N}\to 0$, see appendix A.4 in \cite{Antipin:2014qva}.
In this limit the determinant of the mass matrix (\ref{massmatrix}) vanishes signalling a zero eigenvalue. 
Therefore $\theta_H$ is not physical for zero masses, no barrier is generated and therefore the relaxation cannot be realised. 
The dynamics is such that the singlet $\eta$ settles to a minimum that cancels the $\theta_H$ dependence. 

For small (but non-zero) fermion masses, the formula for the barrier (\ref{barrier}) 
will be valid as long as the fermion mass contribution to $m_\eta$ in (\ref{pionmasses}) is larger than the Yukawa contribution in (\ref{masseta}). 
This corresponds to,
\begin{equation}
m_{L,N}> |y \tilde{y}|  \frac {v^2}{m_{K_2}^2} g_\rho f_\pi
\end{equation}
which is equivalent to the request $\langle \eta \rangle< f_\pi$.

\section{Conclusions}
\label{end}

New dynamics at the TeV scale motivated by naturalness of the Higgs mass appears increasingly unlikely. 
For this reason it is timely to think about new scenarios that provide plausible new physics that
is not already strongly constrained by current experiments but could be tested at LHC run II and beyond.
Among these ``unnatural'' scenarios, the framework of vector-like confinement stands out. New  fermions in a real representation of the SM 
and charged under  a new confining gauge theory, automatically provide electro-weak preserving new physics
for which very weak bounds exist. In particular these models explain without any contrivance why the new physics does not produce new flavour effects and
give controllable effects in precision tests. Their discovery would be analogous to the one of the muon after the electron: 
nobody ordered them but they are completely consistent with all we know. 
These models also provide automatic dark matter candidates that are stable due to the accidental symmetries of the theory in completely  
analogous way as the proton in the SM.

In this note we began the study of models where the quantum numbers of the new fermions allow for Yukawa couplings to the  Higgs.
For simplicity we focused on the most economical choice with a lepton doublet $L$ and a singlet $N$ but our findings are generic. 
The presence of Yukawa couplings makes the phenomenology very rich. Among the composite pions there is always
a state with the same quantum numbers as the elementary Higgs so that these extensions of the SM realise a Two Higgs doublet Model
where one Higgs is elementary and the other is composite. We studied here the case where the lighter Higgs is mostly elementary.
One could also take a different point of view where the light Higgs is composite and the elementary Higgs serves to generate
the appropriate potential. We will explore this scenario further in upcoming work.

The Yukawa couplings introduce new CP violating phases in the theory. This is  similar to split supersymmetry with the difference that now the new fermions 
are confined and the relevant degrees of freedom are the pions. We have shown how the computation of EDMs can be carried out within the chiral lagrangian.
Important effects are obtained for EDMs that could provide the  strongest constraint on this class of models. Some effects in electro-weak 
precision tests and Higgs couplings are also predicted but the size is controlled by the degree of compositeness of the light Higgs that
can be chosen arbitrarily small.

Interestingly the same type of dynamics is relevant for a non-conventional solution of the hierarchy problem introduced in \cite{relaxion}.
In that Ref. a dynamical relaxation of the electro-weak scale was proposed where the Higgs mass is scanned  by 
an axion field and gets trapped to a value much smaller than the cut-off. The dynamics used is precisely the one of the $L+N$ model
studied here. We have shown that the ``relaxion'' mechanism  also works in the region of parameters where all the fermions are confined which is 
different from the original work. This allows more freedom in the choice of the parameters 
and has very different phenomenology that can be tested in future experiments. 

\small

\subsubsection*{Acknowledgments}
The work of OA and MR is supported by the MIUR-FIRB grant RBFR12H1MW. MR would like to thank the CERN theory division for 
support and hospitality. We would like to thank Roberto Franceschini,  Alex Pomarol and Alessandro Strumia for useful discussions and especially 
Diego Becciolini for collaboration at the initial stages of this work. 

\appendix
\section{Vacuum Energy}

In this appendix we review the dependence of the vacuum energy on $\theta_H$ for the general fermion masses, see \cite{wittenchiral,derafael}.

The VEV $\langle U\rangle$  is determined dynamically by minimising the potential: 
\be
\mathcal{L}= (g_\rho f_\pi^3 Tr[MU]+ h.c) +\frac{f_\pi^2}{16}\frac a N \bigg[\ln (\hbox{det}\ U)- \ln (\hbox{det}\ U^\dagger)\bigg]^2
\ee
where $m_{\eta'}^2\approx 3 a/N$ and $\theta_H$  has been rotated to the fermion mass matrix so that $M= e^{i \frac{\theta_H}3} M_0$ with $M_0$
a diagonal matrix with positive entries. 

Treating the Yukawa as a perturbation one can conveniently look for a solution of the form
\begin{equation}
\langle U \rangle=e^{-i \frac{\theta_H}3}\hbox{Diag}\,(e^{i\phi_L},e^{i\phi_L},e^{i\phi_{N}})\,.
\label{Uvev}
\end{equation} 
the potential becomes,
\begin{equation}
V\approx \frac{f_\pi^2} 4 \left( (- 16 \mu_L^2 \cos \phi_L-8  \mu_N^2 \cos \phi_N)+ \frac a N (2\phi_L+\phi_N- \theta_H)^2\right)
\end{equation}
where $\mu_{L,N}^2\approx m_{L,N} g_\rho f_\pi$. The vacuum is determined by the Dashen's equations,
\begin{equation}
4 \mu_L^2 \sin\phi_L = \frac{a}{N} (\theta_H -2 \phi_L- \phi_N)\,,~~~~~~~~~~~~~~~4\mu_N^2 \sin\phi_N = \frac{a}{N} (\theta_H -2 \phi_L- \phi_N)
\end{equation}
The vacuum energy is,
\begin{equation}
V(\theta_H)=\frac{f_\pi^2}4\left( (-16 \mu_L^2 \cos \phi_L - 8 \mu_N^2 \cos \phi_N) + 16\frac N a \mu_N^4 \sin^2 \phi_N\right)
\label{vacenergy}
\end{equation}
evaluated on the solution above.

We can find approximate solutions for $\mu_{L,N}^2\ll a/N$:
\begin{itemize}
\item{$m_N\ll m_L$:
\begin{equation}
\phi_L \ll 1\,,\phi_N \approx \theta_H~~~~~~~ \Longrightarrow~~~~~~~ V(\theta_H)\approx  - 2  g_\rho f_\pi^3\,m_N \cos \theta_H 
\end{equation}}
\item{$m_L\ll m_N$:
\begin{equation}
\phi_N \ll 1\,,\phi_L \approx \frac {\theta_H} 2~~~~~~~ \Longrightarrow~~~~~~~ V(\theta_H)\approx  - 4  g_\rho f_\pi^3\,m_L  \cos \frac{\theta_H}{2} \end{equation}}
\item{$m_L= m_N$:
\begin{equation}
\phi_N=\phi_L\approx \frac {\theta_H} 3~~~~~~~ \Longrightarrow~~~~~~~ V(\theta_H)\approx  -6   g_\rho f_\pi^3\,m  \cos \frac{\theta_H}3
\end{equation}}
In the last case the vacuum energy has a non-analytical behaviour around $\theta_H=\pi$ where the derivative of the vacuum energy is discontinuous. 
\end{itemize}

\end{document}